\documentclass[twocolumn]{webofc}
\usepackage[varg]{txfonts}   
\usepackage{amssymb}
\usepackage{textgreek}
\usepackage{xcolor}
\usepackage{bm}
\usepackage{wasysym}

\begin{document}
\title{The PanEDM Neutron Electric Dipole Moment Experiment at the ILL}
%
%

\author{\firstname{David} \lastname{Wurm}\inst{1} \and
\firstname{Douglas H.} \lastname{Beck}\inst{2} \and
        \firstname{Tim} \lastname{Chupp}\inst{3} \and
        \firstname{Skyler} \lastname{Degenkolb}\inst{4} \thanks{\email{degenkolb@ill.fr}}\and
        \firstname{Katharina} \lastname{Fierlinger}\inst{1} \and
         \firstname{Peter} \lastname{Fierlinger}\inst{1} \and
         \firstname{Hanno} \lastname{Filter}\inst{1} \and
        \firstname{Sergey} \lastname{Ivanov}\inst{5} \and
         \firstname{Christopher} \lastname{Klau}\inst{1} \and
         \firstname{Michael} \lastname{Kreuz}\inst{4} \and
        \firstname{Eddy} \lastname{Lelièvre-Berna}\inst{4} \and
         \firstname{Tobias} \lastname{Lins}\inst{1} \and
         \firstname{Joachim} \lastname{Meichelb\"ock}\inst{1} \and
         \firstname{Thomas} \lastname{Neulinger}\inst{2} \and         
         \firstname{Robert} \lastname{Paddock}\inst{6} \and         
         \firstname{Florian} \lastname{R\"ohrer}\inst{1} \and
        \firstname{Martin} \lastname{Rosner}\inst{1} \and
        \firstname{Anatolii P.} \lastname{Serebrov}\inst{5} \and
        \firstname{Jaideep} \lastname{Taggart Singh}\inst{7} \and
        \firstname{Rainer} \lastname{Stoepler}\inst{1} \and
        \firstname{Stefan} \lastname{Stuiber}\inst{1} \and
        \firstname{Michael} \lastname{Sturm}\inst{1} \and
        \firstname{Bernd} \lastname{Taubenheim}\inst{1} \and
        \firstname{Xavier} \lastname{Tonon}\inst{4} \and
        \firstname{Mark} \lastname{Tucker}\inst{8} \and
        \firstname{Maurits} \lastname{van der Grinten}\inst{8}\and
         \firstname{Oliver} \lastname{Zimmer}\inst{4}}

\institute{Physikdepartment, Technische Universit\"at M\"unchen, Garching, Germany
\and
           Department of Physics, University of Illinois, Urbana IL, USA
\and
			Department of Physics, University of Michigan, Ann Arbor MI, USA
\and
           Institut Laue-Langevin, Grenoble, France
\and
		  Petersburg Nuclear Physics Institute, Gatchina, Russia
\and
	University of Bath, Bath, UK
\and
		 National Superconducting Cyclotron Laboratory and Department of Physics and Astronomy, Michigan State University, East Lansing MI, USA
\and
STFC Rutherford Appleton Laboratory (RAL), Didcot, UK
}

\abstract{
The neutron’s permanent electric dipole moment $d_n$ is constrained to below $3\times10^{-26}  e~\text{cm}$ (90\% C.L.) \cite{BAKER2006,nEDMlimit2015}, by experiments using ultracold neutrons (UCN).
We plan to improve this limit by an order of magnitude or more with PanEDM, the first experiment exploiting the ILL's new UCN source SuperSUN.
SuperSUN is expected to provide a high density of UCN with energies below 80~neV, implying extended statistical reach with respect to existing sources, for experiments that rely on long storage or spin-precession times.
Systematic errors in PanEDM are strongly suppressed by passive magnetic shielding, with magnetic field and gradient drifts at the single fT level.
%
%
A holding-field homogeneity on the order of $10^{-4}$ is achieved in low residual fields, via a high static damping factor and built-in coil system.
No comagnetometer is needed for the first order-of-magnitude improvement in $d_n$, thanks to high magnetic stability and an assortment of sensors outside the UCN storage volumes.
%
%
PanEDM will be commissioned and upgraded in parallel with SuperSUN, to take full advantage of the source's output in each phase.
Commissioning is ongoing in 2019, and a new limit in the mid $10^{-27}  e~\text{cm}$ range should be possible with two full reactor cycles of data in the commissioned apparatus.
}
\maketitle
\section{Introduction}

\label{intro}

The effective Hamiltonian, describing in general the sensitivity of spin-precession experiments to a permanent electric dipole moment (EDM) $d$, is
\begin{equation}
H = - \frac{\bm{\mathrm{S}}}{|S|} \cdot ( \mu \bm{\mathrm{B}} + d \bm{\mathrm{E}}),
\label{eqn:1}
\end{equation}
where $\bm{\mathrm{S}}$ is the total spin operator (other sources of angular momentum are assumed to vanish), and the classical electric and magnetic fields are $\bm{\mathrm{E}}$ and $\bm{\mathrm{B}}$ respectively \cite{PhysRev.112.1642}.
The number $\mu$ is the signed magnitude of the magnetic moment, which like $d$ may be considered as a coupling constant parameterizing the interaction strength with the appropriate field.
The precession frequency $\omega$ is altered by changes in the strength or relative orientation of the fields; this dependence is used to establish a value or bound for $d$ when the other quantities in equation \ref{eqn:1} are known. The EDM may also be viewed as adding or subtracting phase to the usual Larmor precession.

%
%
%
%
%

Experimental searches for a neutron EDM $d_n$ are usually based on coherent magnetic pulse sequences applied to a polarized ensemble, enabling precise determination of $\omega$ by time-domain interferometry. The statistical sensitivity per measurement is then limited to

\begin{equation}
\sigma\left(d_n\right) \gtrsim \frac{\hbar} {2 \alpha |\bm{\mathrm{E}}| T \sqrt{N}},
\label{eqn:2}
\end{equation}
with $T$ the baseline free precession interval, and $N$ the total number of contributing neutron detection events \cite{pendlebury1984search}. The interferometric visibility $\alpha$ is a numerical factor corresponding to the measured ensemble spin polarization.

At a given level of statistical precision, it is also necessary to constrain or eliminate certain ``systematic effects'' that can mimic an EDM and arise from unstable or imperfectly known conditions. Experiments with stored ultracold neutrons (UCN) are very sensitive to spatial gradients and drift in $\bm{\mathrm{B}}$: for $d_n = 1 \times 10^{-28}e~\text{cm}$, a magnetic field drift of 35~aT produces an equivalent change in $\omega$. 
Two UCN storage chambers can be used simultaneously to cancel magnetic field drifts in first order \cite{gatchinaEDM2015}; a key feature of PanEDM is the use of this system in a stable and well-characterized magnetic environment \cite{shieldDamping2015,MSR2014}.
Nearby magnetometers are still needed to correct for residual drifts and gradients, but systematic effects are sufficiently constrained without introducing an atomic ``comagnetometer'' into the neutron storage cells.

\section{The PanEDM apparatus}
\label{apparatus}

\subsection{Central components: two-cell spectrometer}
\label{central_components}
\begin{figure} 
\begin{center}
\includegraphics[width=0.75\columnwidth]{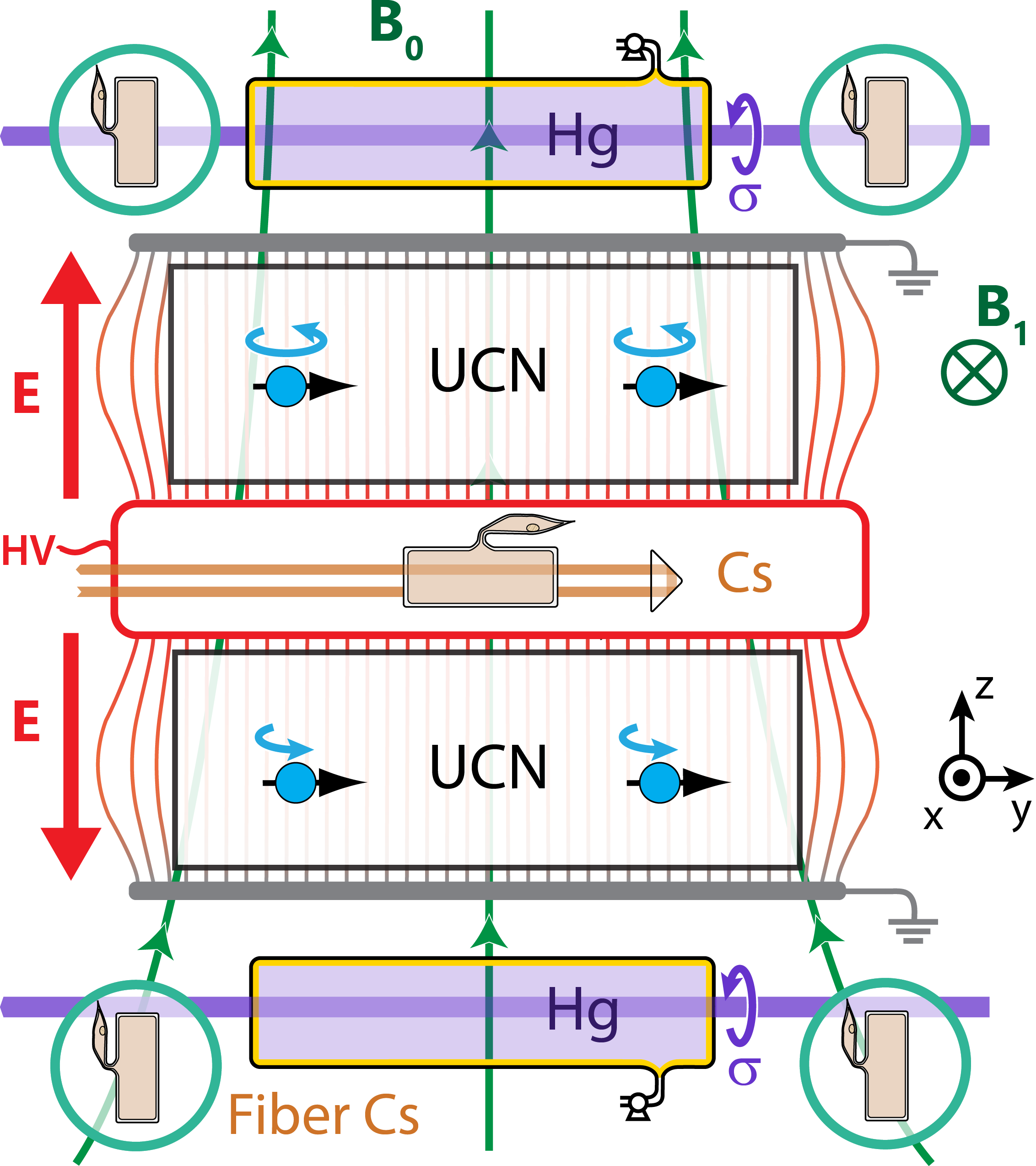}
\caption{Central elements of the PanEDM double-cell spectrometer. Two cylindrical storage cells are filled simultaneously with UCN from a common source, and subjected to a common applied magnetic field $|\bm{\mathrm{B}}_0|\approx 1.3~\mu$T. Equal and antiparallel electric fields with $|\bm{\mathrm{E}}|\approx 2$~MV/m are applied via conductive end-caps, including a common high-voltage electrode in the center. Optical magnetometers based on laser spectroscopy of room-temperature $^{199}$Hg and $^{133}$Cs in cells are arranged around the UCN storage chambers, including Cs sensors with cells inside the hollow central electrode.\label{fig:cellstack}}
\end{center}
\end{figure}

The planned configuration of UCN storage cells is shown in Fig.~\ref{fig:cellstack} without the valves, filling guides, and vacuum chamber. A common central electrode applies high-voltage (HV) of either polarity to both cells at once, producing equal and antiparallel electric fields. The two identical cylindrical cells have quartz insulator rings ($\diameter480\text{~mm}\times94\text{~mm}$) coated with deuterated polyethylene (dPe) \cite{dPE,tito_lanlreport_internal}, and flat electrode end-caps coated with copper or diamond-like carbon. Visibility parameters ${\alpha(T=200~s) > 0.8}$ were demonstrated in storage experiments with test cells at PF2 \cite{ThesisZechlau}. 

This configuration enables simultaneous measurements with $\bm{\mathrm{E}}$ and $\bm{\mathrm{B}}$ parallel and antiparallel. Gradients, drifts, and systematic effects associated with slow magnetic field changes are monitored by optical mercury magnetometers (section \ref{Hg}) above and below the UCN cells. Arrays of higher-bandwidth Cs magnetometers (section \ref{Cs}) monitor higher spatial field modes, and detect transient magnetic fluctuations such as discharges.

The hollow central electrode contains additional Cs magnetometers at high potential, but low electric field. These are physically decoupled from ground and interrogated by free-space laser beams. A leakage-current monitor also sits at high potential (section \ref{HV_monitors}), with optical decoupling for both power and readout.

UCN are filled into the cells through nonmagnetic shutter valves on the ground electrodes. The entire stack sits in a nonmagnetic vacuum chamber made from laminated glass-fiber composite, with quartz optical windows for laser access. Throughgoing tubes at ambient pressure are accessible from outside the magnetic shields, for easy placement of certain magnetometers (or other sensors).
%

%

%
A homogeneous magnetic field $\bm{\mathrm{B}}_0$, initially $1.3~\mu$T, is applied along the $\hat{z}$ axis with a cosine theta coil, split for compensation of vertical residual gradients. The magnetic resonance fields $\bm{\mathrm{B}}_1$ are applied through a similar coil oriented along the $\hat{y}$ axis. Both coils wrap a cylindrical mu-metal (Magnifer${}^\circledR$) shield of 1.4~m diameter and 2.2~m length, which forms the innermost layer of the magnetic shielding described in section \ref{shields}. The coils are designed such that their ``stray'' fields are spatially uniform inside the cylinder, while any remaining gradients are trimmed by end-correction coils \cite{stuiberThesis}.

\subsection{Major infrastructure}
\label{bigstuff}

\begin{figure*}[h]
\centering
\includegraphics[width=1.6\columnwidth]{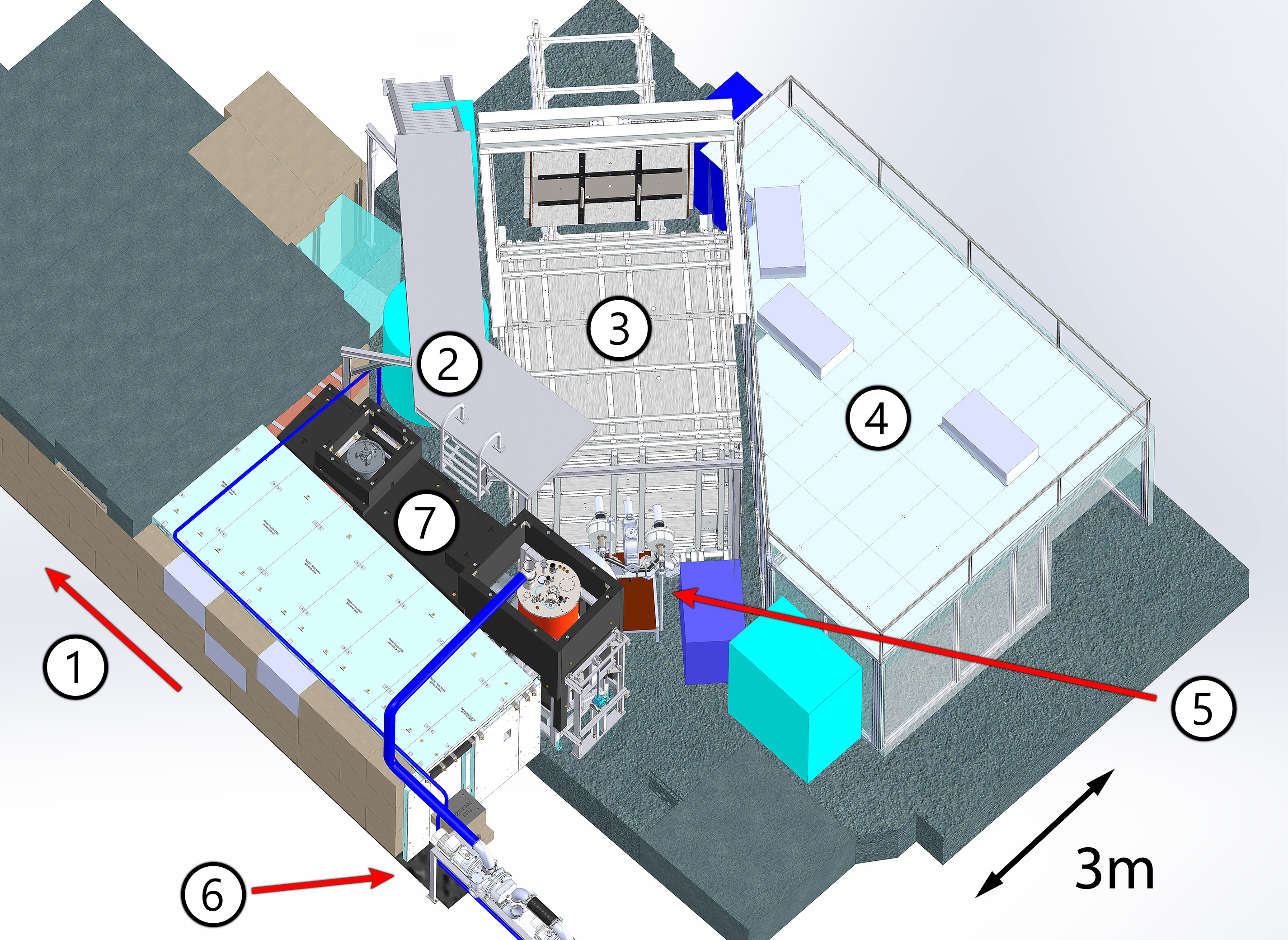}
\caption{SuperSUN/PanEDM experimental area. 1: Towards reactor, 2: service platform above magnetic shields, 3: outer magnetic shield (MSR; detail in Fig.~\ref{fig:shield}), 4: cleanroom, containing Cs and Hg magnetometer control, 5: HV apparatus and UCN optics (detail in Fig.~\ref{fig:guidesystem}), 6: SuperSUN ${}^3$He pump system, 7: SuperSUN.\label{fig:Site}}
\end{figure*}

The experimental zone layout is shown in Fig.~\ref{fig:Site}. The  assembly described above is placed in a configuration of cuboid magnetic shields (section \ref{shields}), consisting of an outer magnetically shielded room (MSR) and a removable insert.
%
%
A cleanroom is installed next to the magnetic shields to house lasers and low-noise electronics, and for handling sensitive components. 
The UCN guiding and detection system (section \ref{guides}), together with HV apparatus, fills the space between the UCN source and magnetic shields. The neutron guides, HV insertion, and vacuum connections enter the shields through $\diameter 130$~mm holes opposite the MSR door.


\subsection{UCN source: SuperSUN}
\label{SuperSUN}
The new ILL source SuperSUN \cite{SuperSUNzimmer2016} will provide UCN at the end-position of beamline H523 \cite{octoguide}; PanEDM will be its first user experiment, and is being commissioned with the source as shown in Fig.~\ref{fig:Site}. SuperSUN is based on the conversion of cold neutrons to UCN by inelastic scattering in isotopically pure superfluid ${}^4$He \cite{golub1977interaction,GOLUB1975133}. The dominant mechanism is single-phonon excitation by neutrons in a narrow range of wavelengths around 8.9~\AA{} \cite{baker2003experimental}. The superfluid is held below 0.6~K to suppress loss due to inverse conversion processes where UCN gain energy from thermal phonons. The saturated maximum UCN density in the source is $\rho = P \tau$, where $P$ is the production rate density (proportional to the flux of 8.9~\AA{} neutrons) and $\tau$ is the storage lifetime for UCN in the converter vessel.
SuperSUN is based on prior work with two prototype sources \cite{zimmerPRL2011,zimmerSUN2016}, of which SUN2
has produced a record \emph{in-situ} density of 220~cm$^{-3}$ with no corrections applied to the extraction measurement.  The spectral maximum is near 80~neV, with half-maximum points at approximately 71~neV and 88~neV measured by time-of-flight (TOF) after 200~s accumulation in the source. The spectrum is observed to shift towards lower energies for longer accumulation. In SuperSUN phase I, UCN will be trapped in the converter by material walls only, and a similar spectrum is expected. The converter volume is 12 liters (three times larger than in SUN2); scaling for this and the brighter cold beam implies a production rate on the order of $10^5$~s$^{-1}$. At saturation, a total of $4 \times 10^6$ stored UCN is predicted ($330~\text{cm}^{-3}$).

In phase II, a magnetic octupole reflector\footnote{currently in production by Elytt Energy, S.L.} will be added to reduce wall losses in the converter for low-field seeking UCN \cite{zimmerGolub2015}. This provides in-situ UCN polarization, while also increasing both $\tau$ and the maximum UCN energy that can be trapped. The predicted \emph{in-situ} stored, polarized UCN population is $2\times 10^7$ ($1670$~cm$^{-3})$.


%
%
%
UCN are extracted into evacuated guides by mechanically opening a cryogenic valve at the end of each accumulation cycle. Each neutron gains 18.5~neV from leaving the superfluid, which is more than compensated by the gravitational potential gained in passing a 28~cm vertical guide section. A thin ($<\!1$~$\mu$m) polypropylene foil with a stainless steel reinforcing grid separates the vacua in the source and EDM chambers, introducing a measured loss of 10\% per transmission.
\subsection{UCN polarization, guiding, and detection}
\label{guides}

\begin{figure}[h]
\begin{center}
\includegraphics[width=0.99\columnwidth]{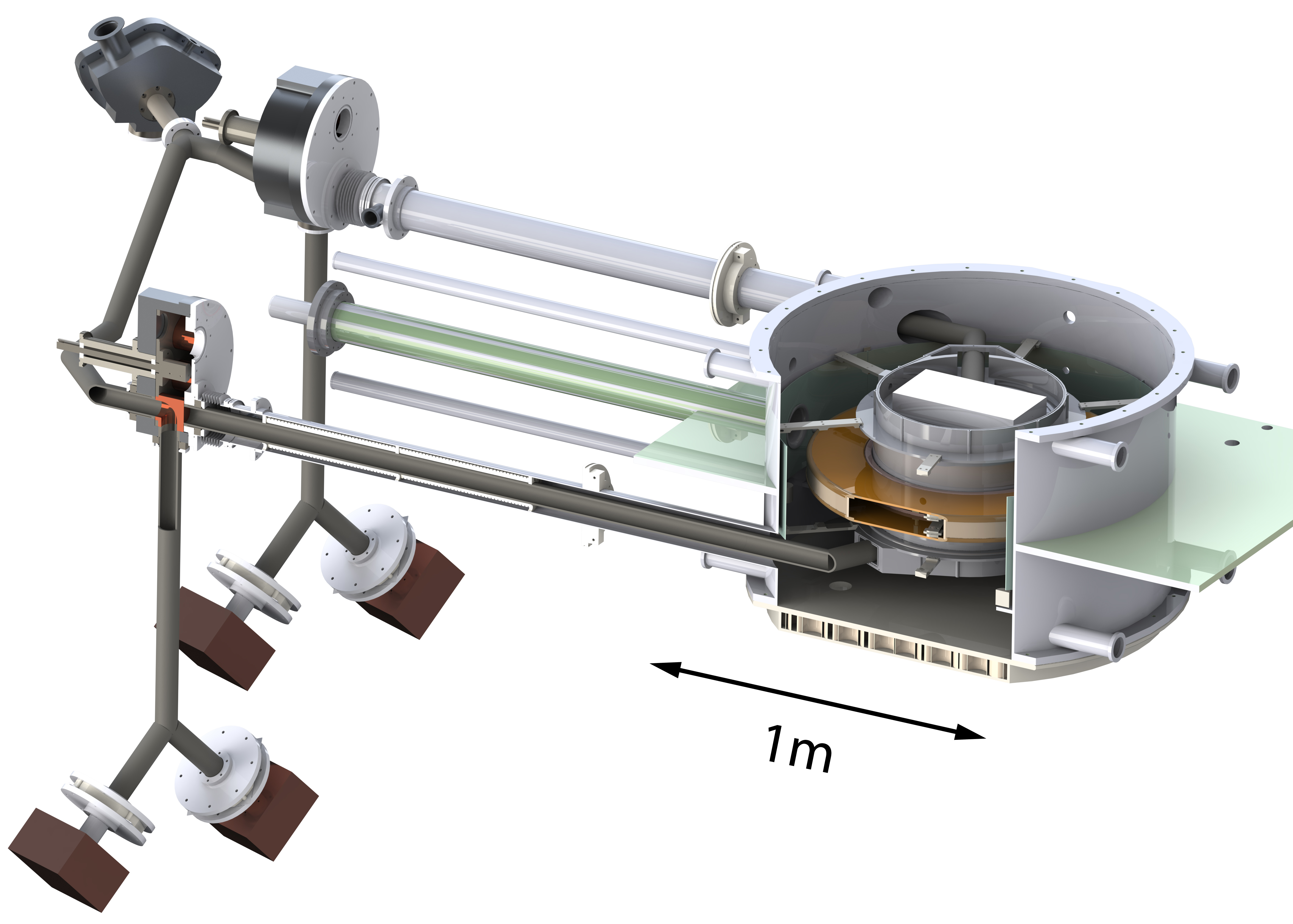}
\caption{UCN guiding and detection system. Guides are glass, with 50~mm inner diameter and nondepolarizing Cu or NiMo coatings. A three-way switch (upper left) allows inclusion of a monitor detector before the polarizer (not shown). Switches following the ``Y'' junction select the filling or emptying path for each cell independently. Each chamber has a dedicated detector system to simultaneously record both spin components from each cell (polarizers shown as white disks). The glass-fiber plastic vacuum chamber is shown in cutaway, also exposing the hollow central electrode and the loading valve for the upper UCN cell.\label{fig:guidesystem}}
\end{center}
\end{figure}

The filling and emptying guides connecting SuperSUN to PanEDM's storage cells and detection system are shown in Fig.~\ref{fig:guidesystem}. The guide substrates are $\diameter 50$~mm glass tubes, with nickel-molybdenum 85/15 (NiMo) or copper coating; these coatings were observed to perform similarly in the relevant velocity range, in polarized TOF measurements at PF2 \cite{hingerl}. The distance between SuperSUN and the outer MSR wall is chosen to minimize the guide volume, while preserving a symmetrical ``Y'' junction. Accounting for dilution and transport losses, the expected initial UCN density in the EDM cells is $3.9$~cm${}^{-3}$ for phase I.

For commissioning, extracted UCN are polarized by a magnetically saturated iron coating on a polished aluminium foil. To reach the phase I target sensitivity, this is to be replaced by a magnet supplying 1.7~T in free space, adequate to polarize UCN up to 100~neV. (Although UCN up to 195~neV may be produced in SuperSUN, those with over 100~neV are not efficiently transported or stored.) Adiabaticity in the guides crossing the MSR wall is maintained by compensated solenoid coils, which need not be switched off during spin-precession periods; this avoids disturbing the shields' magnetic equilibration.

%
%
%
%
After a storage cycle UCN are released from each EDM cell, and follow the filling guides until repositioned switches direct them into a dedicated apparatus for simultaneous detection of both spin states. Two CASCADE detectors\footnote{CASCADE is a trademark by CDT detector technologies} are used for each cell, with an iron-foil polarizer and adiabatic spin-flipper positioned immediately before each one. By alternating which spin-flippers are active, any detector can be made exclusively sensitive to either spin state. Changing this assignment during each measurement helps avoid systematic effects or reduced visibility associated with differences between detectors.

The detection polarizers are placed $113$~cm below the SuperSUN outlet, to ensure that all high-field-seeking UCN have sufficient kinetic energy to pass through. The low-field-seeking spin state is still reflected for even the most energetic UCN, since the dynamic range of an iron-foil polarizer reaches approximately 330~neV.



%
%
%
%
%

\subsection{Electromagnetic shielding}
\label{shields}
\begin{figure}[h]
\begin{center}
\includegraphics[width=0.95\columnwidth]{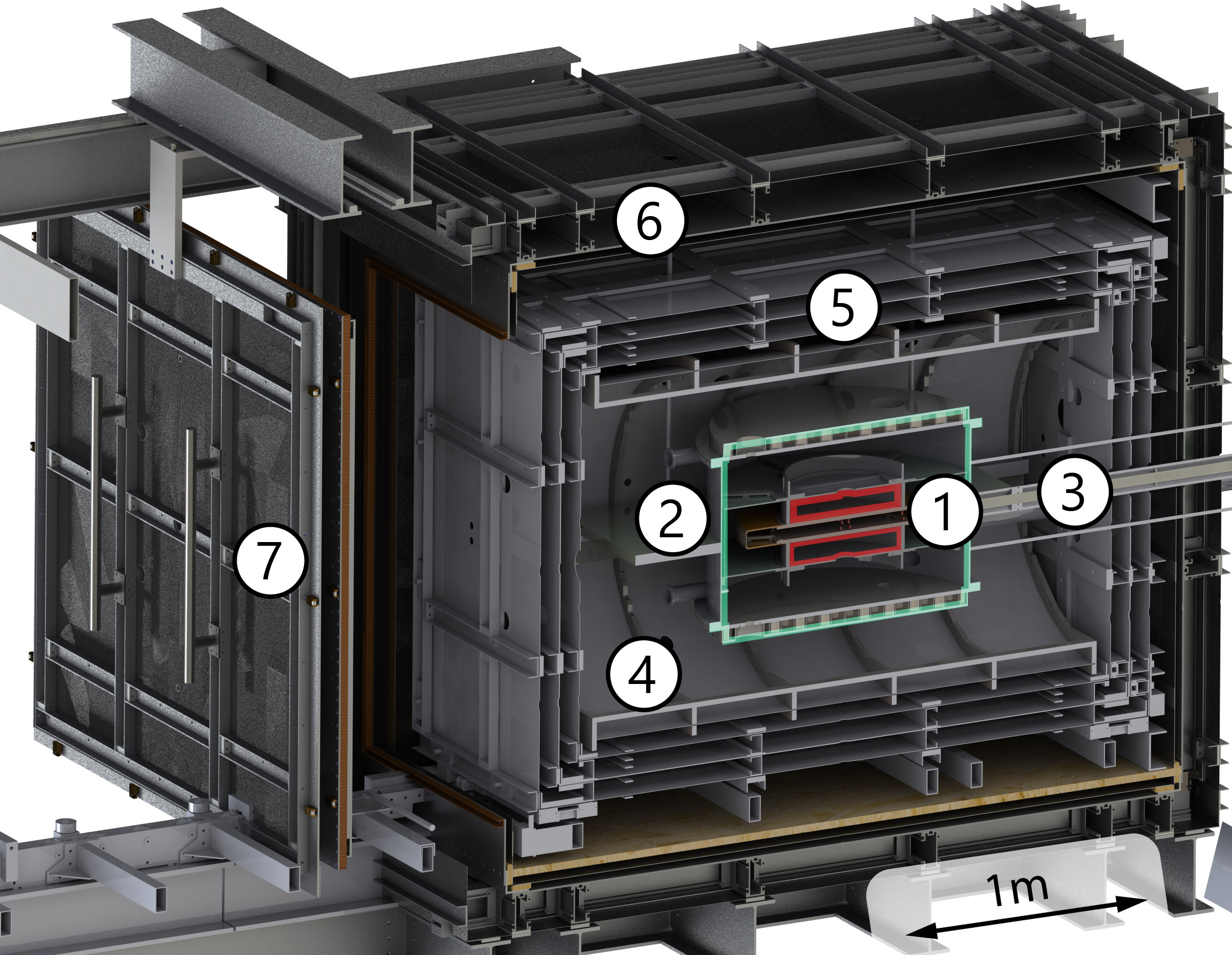}
\caption{PanEDM magnetic shields and central components. 1: UCN cells, 2: vacuum chamber, 3: high-voltage insertion, 4: cylindrical shield and field coils (not shown) for $\bm{\mathrm{B}}_0$ and $\bm{\mathrm{B}}_1$, 5: three-layer inner magnetic shield (Insert), 6: outer magnetic and RF shield (MSR), 7: MSR door \label{fig:shield}}
\end{center}
\end{figure}

%
Magnetic field gradients cause multiple systematic effects, thus stringent control is required over their magnitude \cite{Pendlebury04} and temporal stability \cite{shieldDamping2015}.
The required values for PanEDM's target sensitivity were achieved, using a passive mu-metal shield with a mHz-regime damping factor of $6 \times 10^{6}$ \cite{MSR2014,shieldDamping2015}, shown in Fig.~\ref{fig:shield}.
The relative holding-field homogeneity is $5\times10^{-4}$ over one UCN cell, limited by the mapping device's mechanical tolerance \cite{stuiberThesis}.
The temporal gradient stability of the residual field, measured at the FRM II without the cylindrical shield and field coil, is better than $10$~fT m$^{-1}$s$^{-1}$ \cite{lins2016high}.
%

%
%
%
%
%
%

%
In contrast to degaussing, which is done by forcing the material to have $\bm{\mathrm{H}}=\bm{\mathrm{M}}=0$ in a near-zero total magnetic field $\bm{\mathrm{B}}$, magnetic equilibration is done in the presence of significant internal and external fields like $\bm{\mathrm{B}}_0$ and Earth's field.  
%
%
Successive magnetization and demagnetization minimizes the potential energy of magnetic domains in the shield walls, thus optimizing the residual field inhomogeneity in the desired field configuration.
For sufficiently high static damping factors, the magnetic fields of these domains (and not those from external sources) dominate residual fields inside the shields.
Using ``L''-shaped equilibration coils has reduced the time required for equilibration of one shielding layer from 300~s to 30~s \cite{equilibration_lshape}, implying a considerable reduction in the duty factor associated with reversal of $\bm{\mathrm{B}}_0$.
%
%
%
%

The currents in 43 additional coils can be adjusted to minimize gradients in the presence of $\bm{\mathrm{B}}_0$.\
Quadratic interpolation gives a vertical gradient of ${\partial_z(\bm{\mathrm{B}}_0\cdot\hat{\bm{z}})\approx1.25}$~nT/m over 10~cm, based on a measurement with iterative adjustment of only 14 coils \cite{stuiberThesis}.
This will be reduced to the range of $0.1-0.3$~nT/m by using more coils, implying a geometric-phase-induced false EDM of $1-5 \times 10^{-28}e~\text{cm}$ \cite{EDMreview2018} (see section \ref{GP}).
%
%
Since relatively low currents are used in the correction coils, they contribute neglible noise and produce only percent-level changes in the total field magnitude $|\bm{\mathrm{B}}|$.
%
%
%
%

%
\subsection{Magnetic field sensing}
\label{magnetic sensing}
The double-chamber spectrometer compensates first-order drifts directly, and higher-order drifts can be analyzed through measurement sequences following certain permutations of the HV patterns ($+--+$) and ($-++-$) \cite{gatchinaEDM2015}.
Random fluctuations are also suppressed by repeated measurements, but drifts or offsets that correlate with the sign of $\bm{\mathrm{E}}\cdot\bm{\mathrm{B}}$ can appear as a false EDM if not specifically monitored or prevented.
Therefore a set of additional magnetometers monitors spatiotemporal magnetic field variations during EDM measurements, to directly or indirectly establish limits on the size of such effects.

False EDMs can arise from several distinct sources, including fields due to leakage currents, magnetization due to HV charging currents, $\bm{\mathrm{E}}$-linear geometric phases, imperfect reversal of $\bm{\mathrm{E}}$ and $\bm{\mathrm{B}}_0$, and drifts in the gradient tensor $\bm{\nabla}\bm{\mathrm{B}}$.
Each such effect must be constrained using sensors adapted for its respective source; various sensors are thus needed with differing bandwidth, noise spectra, drift stability, physical size, spatial resolution, and absolute versus relative calibration.
The sensor suite described below is sufficient -- without comagnetometry -- to constrain the total systematic error in $d_n$ from known sources at the $10^{-27}e~\text{cm}$ level.
As-yet-unknown effects could still generate false EDMs via un-monitored channels; this is increasingly a concern at higher experimental sensitivity, and motivates thorough characterization of any process that could cause shifts in the measured precession frequency.

%
%
%

\subsubsection{Mercury magnetometers}
\label{Hg}
Magnetometer cells with Hg vapor at $10^{-4}$~mbar are placed directly above and below the UCN cells; these are the primary means of measuring magnetic gradient stability across the stack.
The cylindrical quartz cells have 60~mm inner diameter and 200~mm length, and internal paraffin antirelaxation coatings.
Their dimensions are chosen to approximately match the gradient-induced transverse relaxation time of precessing $^{199}$Hg with anticipated UCN holding times of $\sim\!250$~s.
%
%
In test measurements with a 40~mm diameter cell (limited by wall depolarization), a sensitivity of 12~fT was achieved in 100~s of integration.

%
The 6${}^{1}$S$_{0} \rightarrow 6 {}^{3}$P$_{1}$ intercombination transition at 253.7~nm is sensitive to nuclear spin orientation through hyperfine coupling in the I=$\frac{1}{2}$ isotope $^{199}$Hg, and is used to probe the instantaneous $\hat{\bm{y}}$ spin projection of the precessing ensemble.
%
%
This transition is also exploited for spin-polarization and to initiate precession, using transverse optical pumping \cite{GREENdischarge,BerndThesis} with chopped laser beams.
Light for the pump and probe beams is produced by twice doubling the frequency of a single 1015~nm diode laser.
The laser's fundamental frequency is locked at the value where the 253.7~nm light shift $\nu_{\text{Hg}}$ vanishes, using Doppler-free DAVLL/DFDL \cite{cheron1994laser, Corwin_DAVLL, DFDL} in a separate Hg cell.
The laser power is independently stabilized by an intensity lock, further reducing the influence of $\nu_{\text{Hg}}$ on the measured precession frequency.
This effect, which contributed $0.8 \times 10^{-27}e~\text{cm}$ in reference \cite{nEDMlimit2015}, does not directly enter PanEDM's sensitivity.

The chopped pump beams enter the magnetic shields in free space, and enter the vacuum chamber through antireflection-coated windows before reaching the cells.
Spin precession is initiated during the preparation time for each storage cycle, and the $m_I = \pm\frac{1}{2}$ populations oscillate due to precession at the frequency $7.59~\text{Hz}/\mu\text{T} \cdot |\bm{\mathrm{B}}|$.
This frequency is imprinted as an intensity modulation on the circularly polarized probe beams, which cross the cells and are detected outside the magnetic shields by photodiodes with trans-impedance amplifiers.
The 0.5 $\mu$W probe beams are sufficiently weak to neglect perturbations to the oscillating populations by optical pumping.
%

Due to the large gradiometer baseline between the top and bottom Hg cells, a 4~fT differential sensitivity is adequate to rule out false effects arising from correlations of $\bm{\nabla}\bm{\mathrm{B}}$ with the HV polarity at the $10^{-27}e~\text{cm}$ level.
Operation of the two Hg sensors with a common light source further reduces the possibility for differential drift.

\subsubsection{Cesium magnetometers}
\label{Cs}
All-optical Cs magnetometers are used to monitor magnetic field changes below kHz frequencies, at multiple locations around the UCN cells. 
They are intrinsically less stable than the Hg magnetometers, and thus unsuitable for absolute measurements, but their larger bandwidth and shorter integration time make them appropriate for monitoring transient magnetic effects. 
%
%
Their design is based on that of Budker \emph{et al.}~\cite{NMORreview}, and adapted for low-drift operation with longer integration times.

The sensor heads contain no coils or other metallic components, and all optical components are fixed in a 3D-printed holding structure for mechanical rigidity.
The core element is a cylindrical paraffin-coated Cs vapor cell, with 10~mm outer diameter and 30~mm length.
Each cell is optically pumped and probed via the $852$~nm D$_{2}$ transition using linearly polarized laser light, either from a free-space beam or coupled into the sensor head with an optical fiber.
All sensors use amplitude-modulated nonlinear magneto-optical rotation (AM-NMOR) \cite{NMORreview}, where the pump beam generates circular birefringence and dichroism by redistributing the thermal populations of hyperfine levels in the electronic ground state.
This modifies the polarization and intensity of the transmitted probe beam, which are further modulated as the polarized ensemble precesses in a magnetic field. 
%
%
By pulsing the pump beam at different frequencies, a Lorentzian "forced-oscillation" curve peaked at the precession frequency can be extracted with phase-sensitive detection.
The sensors also operate in a second mode using free precession decay, providing an important consistency check when performed back-to-back with forced-oscillation measurements. 
The cells have transverse coherence times in the range of 30-50~ms, and typical sensitivity below $150$~fT after 250~s of integration.

%

%
%

The fiberized sensors are arranged around the UCN cells to enable reconstruction of high-order spatial field modes \cite{lins2016high}, while free-space sensors are placed inside the hollow HV electrode at low electric field.
The in-electrode sensors provide sensitivity to nearby magnetic dipoles (e.g., ones arising from charging currents) and transient fluctuations in the heart of the experiment, without compromising the local field homogeneity or providing a new path for leakage currents.

%
%
%

\subsubsection{Absence of comagnetometer}
Modern neutron EDM experiments typically rely on nuclear-spin comagnetometers that occupy the same volume as the UCN, to compensate for magnetic field drifts \cite{ramsey1984feasibility, lamoreaux1989electric, BAKER2006, schmidt2016quest, EDMreview2018}. 
This is evidently critical in single-cell experiments; in two-cell spectrometers, common field drifts are automatically corrected within each measurement and drift of $\bm{\nabla\mathrm{B}}$ dominates instead.
Without comagnetometry the required gradient stability and homogeneity are significantly more stringent, but if sufficiently stable magnetic conditions can be provided, other key limitations are relaxed or eliminated.
Interestingly, the shielding factor for gradients in the PanEDM MSR is observed to exceed the corresponding shielding factor for homogeneous fields \cite{MSR2014}.

False EDM effects in atomic gases are avoided by placing all atomic magnetometers in regions of low electric field.
%
%
For example, $\bm{\mathrm{E}}$-linear geometric phase shifts in comagnetometers are typically much larger than for UCN, due to 20-40 times larger velocities; the need to correct for this effect disappears in regions where $|\bm{\mathrm{E}}|\approx 0$.
Eliminating complex apparatus for loading comagnetometer gas directly into UCN cells also eliminates any magnetic field inhomogeneities due to these components, and provides a significant conceptual simplification of the apparatus.

Additionally, longer holding times and higher electric fields are possible in the absence of low-pressure atomic gases; this increases a single run's statistical reach by more than $50\%$.
Deterioration of the storage environment for polarized UCN (associated with the presence of comagnetometer gas \cite{GREENdischarge}) is also avoided, and the need for periodic discharge cleaning removed.
%
%
Coatings, etc., can be selected for their UCN properties without compromising for compatibility with a comagnetometer, and analysis is somewhat simplified by eliminating direct interactions between the UCN and comagnetometer atoms.

The cost of eliminating the comagnetometer is reduced knowledge of the spatiotemporal magnetic field structure inside the UCN cells.
It is thus essential to optimize the performance and placement of external magnetometers, and to combine data from different sensors to constrain any gradient drifts or false EDM effects that would normally be treated by comagnetometry.
Gradient drifts can be extracted from combinations of Hg and Cs signals, while the in-electrode sensors are used to detect magnetization or fluctuating fields near the spectrometer's center. 

\section{Statistical sensitivity}
\label{sec:sensitivity}
PanEDM aims to reach a sensitivity of order $10^{-27}e~\text{cm}$ in 100 days with SuperSUN phase I, followed by major improvements with SuperSUN phase II, as shown in Table \ref{table:limits}.
Electric fields of $2$~MV/m, and values of 250~s for the precession interval and transverse coherence time (including all relaxation and loss mechanisms), are assumed throughout.
SuperSUN's saturated UCN density is estimated as described in section \ref{SuperSUN}; the 400~s duty cycle optimizing $T\sqrt{N}$ for PanEDM does not allow it to reach saturation, so we have reduced the estimated UCN density for each run by $50\%$.
The estimated initial density in the cells includes loss factors for volume dilution (0.19) and transport efficiency (0.25). A loss factor for polarization (0.50) is assumed for phase I only, with the new polarizer. A factor for emptying efficiency (0.40) is also included.
%


The transport factor for the filling and emptying guides is uncertain due to strong dependence on details such as gap sizes and surface quality; the final implementation can only be tested with UCN from SuperSUN.
A detector efficiency near 100\% is assumed, with a visibility ratio of 0.85 for polarization analysis.

\begin{table}
\centering
\caption{Estimated statistical sensitivity, assuming continuous operation with a 400~s repetition period (250~s free precession, with 150~s preparation/detection). One reactor cycle $\approx$ 50 days.}
\label{table:limits}       
\begin{tabular}{l r r}\hline\hline
{\bf SuperSUN} &  {\bf Phase I} & {\bf Phase II}  \\ \hline
Saturated source & & \\
density [cm$^{-3}$]& $330$ & $1670$\\
Diluted density [cm$^{-3}$]& $63 $ & $318$ \\
Density in cells [cm$^{-3}$] & 3.9 & 40\\
\multicolumn{3}{l}{\textbf{PanEDM Sensitivity} [$1 \sigma $, $e~\text{cm}$]}\\ \hline \\[-1.0em]
Per run & $5.5 \times 10^{-25} $&$1.2 \times 10^{-25}$ \\
Per day & $3.8 \times 10^{-26} $&$7.9 \times 10^{-27}$ \\
Per 100 days  & $3.8 \times 10^{-27} $&$7.9 \times 10^{-28}$ \\
   \hline\hline
\end{tabular}
\medskip
\end{table}

\section{Systematic errors and uncertainties}
Many systematic effects are known from previous work \cite{BAKER2006,nEDMlimit2015,EDMreview2018,Pendlebury04}, and others could appear soon \cite{abel2018magnetic,nonExtensive}.
The strategy here, motivated by both the absence of comagnetometry and the prospect for a near-term increase in statistical reach with SuperSUN phase II, is to investigate and characterize all magnetic factors as thoroughly as possible.
%
%

\subsection{High voltage and leakage current monitoring}
\label{HV_monitors}
The central electrode is charged with $\pm 200$~kV, to produce antiparallel electric fields of $2$~MV/m in the two cells. 
A leakage current from this electrode to ground could cause a magnetic field proportional to $\bm{\mathrm{E}}$, which could in turn appear as a false EDM if not corrected in analysis. 
A  50~pA current flowing in a single loop around one cell implies a false EDM of $10^{-28}e~\text{cm}$.
To place experimental upper limits on such errors, a current monitor is built into the HV feed in such a way that all leakage currents flowing near the UCN cells must pass through it.
It floats at high potential, and is powered by a 20~W infrared laser using photocells.
The data are read out by frequency-encoding a blinking LED of a different wavelength; the device has a demonstrated resolution at ground of 500~fA in a 1~Hz bandwidth.
A higher-bandwidth voltage monitor will also complement the free-space Cs sensors, for recording sparks or other transients.

%
%

\subsection{Geometric phases}
\label{GP}
A special class of systematic error can arise from "geometric" or nondynamical phase shifts \cite{Pendlebury04}.
These can be viewed in the framework of Berry's phase \cite{berry1984quantal}, in that each particle's wavefunction depends on parameters such as the magnetic field components $(B_x,B_y,B_z)$, which vary in the particle's rest frame during a measurement.
When the wavefunction in this parameter space follows a closed trajectory that subtends a finite solid angle at a degeneracy point such as $|\bm{\mathrm{B}}| = 0$, a constant phase proportional to that solid angle is added to the dynamical phase of the wavefunction.
%
%
This added phase mimics an EDM frequency-shift if it varies linearly with $\bm{\mathrm{E}}$; this does not occur directly from motional fields of the form $ \bm{\mathrm{E}} \times  \bm{\mathrm{v}} /c^2$, since these add in quadrature to $|\bm{\mathrm{B}}|$.
However, an important false EDM can appear when in addition to motional fields, a gradient exists along the direction of $\bm{\mathrm{B}}_0$, implying finite transverse fields of the form $-( \bm{\mathrm{r}}/2)\partial_z(\bm{\mathrm{B}}_0 \cdot \hat{\bm{z}})$.
``Cross-terms" between the motional and gradient fields then generate a frequency shift that is truly linear in $\bm{\mathrm{E}}$, and must therefore be measured or suppressed to distinguish it from any true EDM. 
For PanEDM a global gradient larger than 0.3~nT/m, or a local gradient caused by a dipole producing 2~pT at 3~cm distance, would cause a false EDM of order $10^{-28}e~\text{cm}$. 

%

\subsection{Other systematics}
Systematic effects can also arise from magnetic field distortions other than vertical gradients.
%
%
The HV feedthrough connects to the HV electrode on the central axis, which could lead to a shift in $\bm{\mathrm{B}}$ if mechanical misalignments occur together with leakage currents.
Second-order effects can also arise from AC magnetic fields due to HV ripple. 
Asymmetries in the HV supply and return can also contribute to frequency shifts.
%
%

A systematic error arising from different UCN energy spectra in the upper and lower cells could be addressed by energy-dependent analysis, such as the spin-echo technique of Afach \emph{et al.}\ \cite{psi_paper_spin_echo}. 
The UCN transport system has been designed with input from simulations \cite{geant4}, to avoid ordered motion inside the cells which could enhance a geometric phase asymmetry.
Apart from the UCN shutter, no mechanical actuators are placed near the cells in order to mitigate the risk of magnetic contamination or irreproducible gradients due to moving components.
%
%
%
Although not yet known to produce frequency shifts, non-ergodic evolution of spin distributions in traps \cite{nonExtensive} is a possible new source of systematic errors; for distributions with antisymmetric higher moments this could also influence known effects.
\section{Outlook}
The PanEDM apparatus is currently being assembled at the ILL, with first UCN measurements expected in 2019.
These will serve as a basis for further commissioning, with first EDM runs beginning in late 2019 or 2020.
At least two full reactor cycles (100 days) of EDM data will be required to achieve a statistical sensitivity in the mid $10^{-27}e~\text{cm}$ range, with UCN from SuperSUN phase I.
The polarized, broader energy spectrum resulting from magnetic storage in SuperSUN phase II (commissioned in 2020 or later) could extend the statistical reach of PanEDM to below $10^{-27}e~\text{cm}$.
In order to reduce systematic errors to a comparable level, the PanEDM apparatus will be successively upgraded on a schedule synchronized with work on SuperSUN.

%

\bibliography{MT-Bib-SMDr2}
\end{document}